\documentclass[sigconf,nonacm, manuscript]{acmart}
\AtBeginDocument{%
  }

\setcopyright{acmlicensed}
\copyrightyear{2026}
\acmYear{2026}
\acmDOI{XXXXXXX.XXXXXXX} 
\acmConference[UbiComp/ISWC Workshop '26]{Wearable Design in Games and
  Entertainment --- ACM UbiComp/ISWC Workshop}{October 11--12,
  2026}{Shanghai, China}
\acmISBN{978-1-4503-XXXX-X/2026/10} 

\usepackage{booktabs}
\usepackage{graphicx}

\begin{document}

\title{ItemCompass: Real-Time Edge-AI Wearable Vision for
       Tag-Free Domestic Object Retrieval}

\begin{teaserfigure}
\centering
  \includegraphics[width=.8\textwidth]{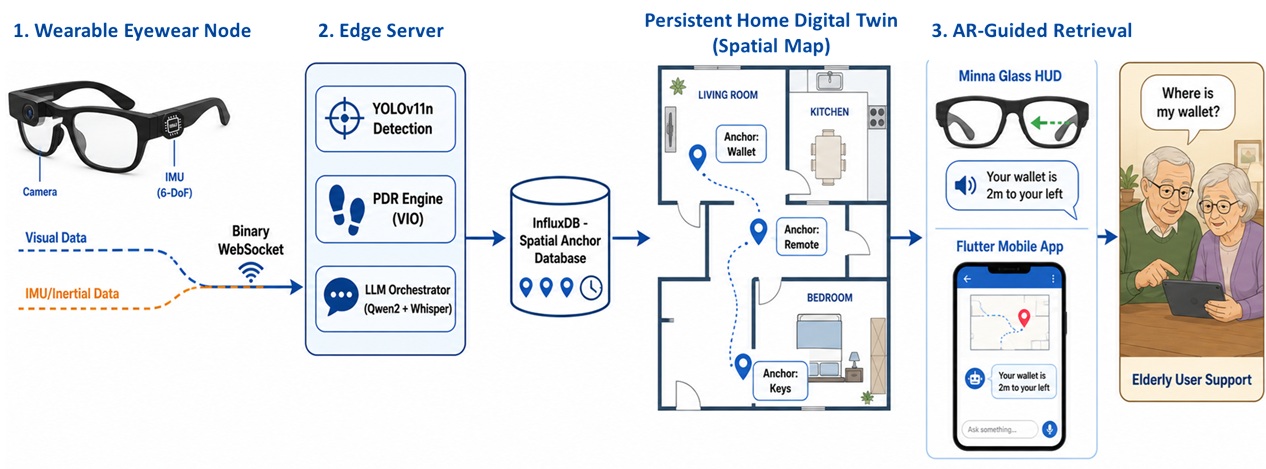}
\caption{ItemCompass streams wearable visual and inertial data to an edge server,
         which builds a persistent home digital twin and delivers an AR-guided item
         retrieval for users (especially the elderly) via HUD and mobile app.}
  \Description{Diagram showing a four-stage system pipeline: smart glasses with camera and IMU sensors send data to an edge server for object detection, motion tracking, and voice processing; results are stored in a spatial map database of a home; the system then guides users to lost items via smart-glasses display and a mobile app, illustrated with an elderly couple asking "Where is my wallet?"}
  \label{fig:teaser}
\end{teaserfigure}

\author{Tsz Lok Cheung}
\orcid{0009-0002-8227-1180} 
\affiliation{%
  \institution{The Hong Kong Polytechnic University}
  \city{Hong Kong}
  \country{China}}
\email{tlcheung2003@gmail.com}

\author{Jeff Kai-Tai Tang}
\orcid{0000-0002-5627-1630} 
\affiliation{%
  \institution{The Hong Kong Polytechnic University}
  \city{Hong Kong}
  \country{China}}
\email{kai-tai-jeff.tang@polyu.edu.hk}
\correspondingauthor

\author{Anthony Kong}
\orcid{0000-0002-8918-5616} 
\affiliation{%
  \institution{The Hong Kong Polytechnic University}
  \city{Hong Kong}
  \country{China}}
\email{anthony-pui-keung.kong@polyu.edu.hk}

\author{Zackary Ping-Tat Sin}
\orcid{0000-0002-8377-0216} 
\affiliation{%
  \institution{The Hong Kong Polytechnic University}
  \city{Hong Kong}
  \country{China}}
\email{zackary-p-t.sin@polyu.edu.hk}

\renewcommand{\shortauthors}{Cheung et al.}

\begin{abstract}
Misplacing household objects disrupts daily routines, especially for elderly users.
ItemCompass addresses this with a real-time, tag-free wearable AR system that builds
a persistent digital twin of the home for natural language object retrieval. A custom
eyewear node streams visual and inertial data to an edge server, where YOLOv11n
detection, Siamese re-identification, and a Pedestrian Dead Reckoning engine log
object locations as Spatial Anchors in InfluxDB. Voice queries are resolved by a
local Faster-Whisper and Qwen2 LLM pipeline, delivering concise spoken and AR
guidance on a heads-up display. Evaluation shows 0.40 m localization drift, 97.5\%
step precision, and mAP@0.5 of 0.874, with users rating Accuracy and Trust at
4.6/5.0. The system highlights always-on sensing, conversational spatial memory, and
privacy-by-design for wearable computing.
\end{abstract}



\maketitle

\section{Introduction}
People routinely misplace everyday objects, and the search that follows disrupts
daily life, acutely so for elderly users prone to spatial disorientation. Wearable
eyewear, with an always-on camera and IMU following the user's gaze, offers a
natural way to passively track where things go without requiring any deliberate
scanning or tagging step. This positions wearable vision as a
compelling substrate for the workshop's focus on lightweight, always-on sensing that
feels like a cognitive extension rather than a conscious tool.

The scale of the problem is substantial. People in the UK misplace approximately 7.9
billion items per year~\cite{DirectLine2020}, and nearly half of Spaniards lose an
item at home weekly, accumulating over six months of search time across a
lifetime~\cite{IKEA2024}. A local survey of 63 Hong Kong residents was conducted as part
of this project found that 42.9\% of respondents lose items weekly and 28.6\% lose
items daily, with remote controls, Octopus cards, bottles, ID cards, mice,
earphones, and wallets were identified as the top seven most frequently misplaced items.

Yet existing solutions fall short in the same way: none combine tag-free tracking
with persistent, queryable spatial memory. BLE trackers such as Apple AirTag achieve
UWB-precision localization~\cite{Kearns2024} but requires physical attachment to
every tracked item. Image-matching systems such as LostNet classify lost items from
photos~\cite{Zhou2024} but offer no real-time spatial guidance. Neither can answer
the simple spoken question that matters most: ``Where did I last see my wallet?''

ItemCompass closes this gap with a three-part wearable edge-AI pipeline: (1)~a
commodity-hardware eyewear node with bimodal visual-inertial streaming that performs
continuous odometry without SLAM; (2)~a spatial anchoring architecture that logs
object detections as timestamped, deduplicated Spatial Anchors with Siamese
re-identification for personalized naming; and (3)~a voice-activated retrieval loop
that delivers landmark-referenced AR guidance at conversational latency, protected by
a hardware-level Privacy Shield in sensitive spaces.

\section{Related Work}
BLE trackers such as Tile and Apple AirTag dominate commercial item-finding, using
BLE proximity~\cite{Pace2023} or UWB precision~\cite{Kearns2024}, but both require
physical attachment to every tracked object. Tag-free alternatives such as LostNet
perform image-matching classification via MobileNetV2 with attention and perceptual
hashing~\cite{Zhou2024}. Neither category provides voice interaction or persistent
spatial guidance. ItemCompass instead couples continuous wearable vision with
inertial odometry to build a queryable 3-D spatial memory of the entire home.

SLAM-based AR tracking suffers from drift and is computationally prohibitive on edge
hardware~\cite{Hu2025}, and commercial smart glasses impose tethering or
subscription-locked APIs~\cite{Rokid2025,Maniar2025}. ItemCompass replaces SLAM with
a lightweight PDR engine, enabling persistent spatial memory on commodity
microcontroller-class hardware. LLM voice interfaces are an emerging wearable
interaction layer~\cite{Maniar2025}. MemPal demonstrates LLM-assisted retrieval for
older adults, which ItemCompass extends with fully local inference and a bilingual
(English/Cantonese) pipeline. Prior AR game-design research further shows that
narratively framing physical space, rather than treating it as a neutral backdrop,
measurably improves engagement~\cite{Li2017}, which is a principle ItemCompass applies by
treating the home as a persistent, narratively addressable spatial memory.

\section{System Design}
\begin{figure}[t]
  \centering
  \includegraphics[width=\linewidth]{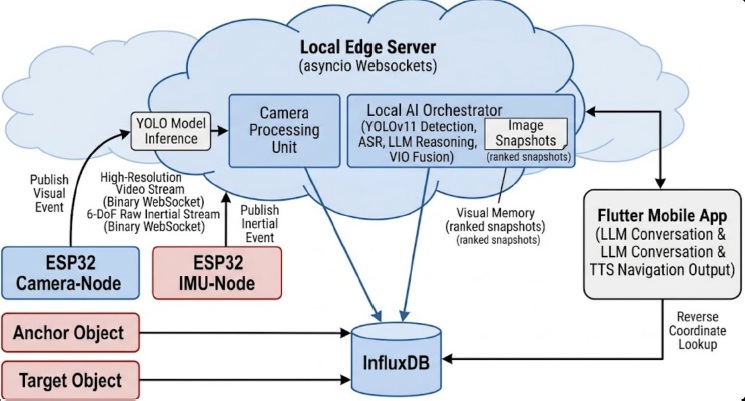}
  \caption{System architecture of ItemCompass: wearable sensing, edge AI, spatial memory, and AR guidance.}
  \Description{Block diagram of the ItemCompass system architecture,
               illustrating the data flow from the wearable AR glasses
               through the binary WebSocket link to the edge server
               AI stack (YOLOv11n, PDR, LLM) and the InfluxDB
               spatial database, with guidance output rendered on the
               Minna Glass HUD and Flutter app.}
  \label{fig:architecture}
\end{figure}

ItemCompass consists of a wearable sensing node and four tightly coupled server-side
subsystems that form a closed sensing-reasoning-guidance loop (Figure~\ref{fig:teaser}).
Figure~\ref{fig:architecture} illustrates how these subsystems interlock, tracing the
full data path from raw sensor capture on the eyewear node to spoken guidance on the HUD.

\subsection{Wearable Compute Node}
The hardware centers on a Seeed Studio XIAO ESP32-S3 Sense node equipped with an
OV3660 CMOS camera (3\,MP, 90$^\circ$ FOV) and an ICM-42686-P 6-axis
IMU~\cite{TDKInvenSense2023}, mounted on a custom 3D-printed eyewear chassis
fabricated with guidance from the PolyU Industrial Centre. An initial iteration
using the ESP32-S3 EYE was abandoned because its onboard QMA7981 IMU provided only
3-axis acceleration is insufficient for stable heading estimation during
turns. The OV2640 camera was similarly replaced: its narrow
FOV caused object tracking to fail during rapid head movements. The MAX98357A I2S
amplifier~\cite{Maxim2020} provides local audio feedback. A custom PCB layout
integrates all components to minimize chassis volume for practical all-day wear.
A critical power-management issue arose during development: simultaneous
initialization of the OV3660 LDO regulators and the MAX98357A amplifier triggered
brown-out reboot loops under USB~5\,V supply. The firmware now duty-cycles the
camera interface to preserve safety-critical IMU fall detection and I2S audio paths.

\subsection{Bimodal Binary WebSocket Protocol}
To minimize latency, the node streams two data types over a single WebSocket connection
using single-byte header multiplexing: \texttt{`V'} for JPEG video frames and
\texttt{`I'} for 12-byte raw inertial packets (Ax, Ay, Az, Gx, Gy, Gz) sampled at
25\,Hz. A 50\,ms throttled transmission loop prevents WiFi stack keepalive errors
under high-volume bursts, achieving a stable sub-100\,ms round-trip latency. The
server uses Python \texttt{asyncio} to process the two streams concurrently without
mutual blocking.

\subsection{Pedestrian Dead Reckoning (PDR) Engine}
The spatial backbone is a custom Visual-Inertial Odometry (VIO) engine that maintains
a 3-D Cartesian coordinate \(\sigma = \{x, y, z, \theta\}\) relative to a fixed
origin. Step detection follows a three-stage pipeline:

\begin{enumerate}
  \item \textbf{Norm calculation}: Acceleration magnitude is computed as
    \(\text{mag} = \sqrt{A_x^2 + A_y^2 + A_z^2}\) so that step
    detection is invariant to eyewear orientation~\cite{Kasebzadeh2016}.
  \item \textbf{Low-pass filtering \& peak detection}: A low-pass filter retains
    walking-frequency content (1--3\,Hz) and suppresses head-jitter noise; peaks
    above a 1.08\,G threshold are candidate steps.
  \item \textbf{Thresholding \& debouncing}: A 450\,ms debounce interval prevents
    double-counting secondary oscillations from a single footfall.
\end{enumerate}

To eliminate ``ghost steps'' caused by head tilts alone, a \emph{Bimodal Validation
Gate} requires both (a)~acceleration magnitude exceeding 1.08\,G \emph{and}
(b)~vertical optical flow \(|\phi_{dy}| > 1.1\,\text{px}\) confirming physical
translation. Stride length is estimated as
\(\text{Stride} = |\phi_{dy}| \times 0.90\), calibrated for 1.6\,m eye-level mounting.
Heading is maintained by a complementary filter fusing the gyroscope Gz with the horizontal
optical flow \(\phi_{dx}\), integrating yaw only when angular velocity exceeds
18$^\circ$/s (``Dead-zone'' heading lock) to isolate body turns from incidental head
movements. When a known Anchor Object (e.g., refrigerator, television) is detected by
YOLOv11n, the engine resets accumulated positional drift using the visual landmark,
providing long-range drift correction without SLAM.

\subsection{Object Detection and Spatial Memory}
A YOLOv11n model fine-tuned with mosaic augmentation on the target object hierarchy
performs continuous inference at sub-15\,ms per frame. Detected objects are passed
through a Siamese re-identification layer: visual embeddings are compared against
user-enrolled anchor embeddings; a cosine-similarity score above 0.85 triggers
personalized naming (e.g., \textit{``Your Wallet''}). A confidence gate of 0.50
filters uncertain detections before any database write.

Confirmed detections are logged as Spatial Anchors in InfluxDB with object label,
category (Target or Anchor), 3-D coordinates, and timestamp. A 0.5\,m Distance Filter
suppresses redundant writes; snapshots are ranked by a decay heuristic
\(\text{score} = \text{confidence} / (\text{age}+1)\), ensuring the highest-certainty
recent frame is served on retrieval.

A \textbf{Privacy Shield} is hard-coded into the pipeline: if YOLOv11n detects a
toilet zone, the server immediately transmits a binary \texttt{`T'} signal to the node,
which plays a local \texttt{alert.mp3} from LittleFS and suspends all server-side
logging, bypassing mobile-application latency for instantaneous enforcement.

\subsection{Voice-Activated AR Retrieval}
Voice interaction is initiated via a dedicated query button on the companion Flutter
mobile application, where spoken audio is transcribed by a local Faster-Whisper ASR
engine. The transcribed query is forwarded to a Qwen2 LLM that performs
retrieval-augmented generation (RAG) over the InfluxDB spatial history and translates
raw coordinates into landmark-referenced instructions (e.g., \textit{``Your Octopus
card is near the television, 2\,m to your left''}). Output is constrained to 10
words for cognitive clarity. Guidance is delivered simultaneously as directional arrows
on the Minna Glass HUD and via Edge-TTS voice output through the mobile application,
supporting both English and Cantonese.

\subsection{System Algorithm}
ItemCompass runs three concurrent operational layers that synchronize raw hardware
data with high-level AI reasoning. \textit{Spatial Awareness} continuously receives
12-byte IMU packets from the node. When the acceleration magnitude exceeds 1.28\,G and
at least 0.65\,s has elapsed since the last step, optical flow is computed from the
current frame, and if vertical flow exceeds a small threshold, the stride length and
position are updated. Heading is fused from the gyroscope and the horizontal flow on every
packet. \textit{Memory and Privacy} runs on every received frame. Detected objects
in a privacy zone (e.g., toilet, bathroom) trigger an immediate local alert and
suspend server-side logging. Otherwise, objects matching an enrolled anchor
embedding above 0.85 cosine similarity are personalized, and any detection with
more than 0.5\,m of motion since the last write is saved to the Spatial Database
together with the current pose. \textit{Retrieval} is triggered on app button
press. The spoken query is transcribed by Whisper, matched against the Spatial
Database, reasoned over by Qwen2 to produce a response of at most ten words, and
delivered simultaneously as HUD text and speech.

\section{Evaluation}

\subsection{Localization Performance}
Five locomotion trials validated the PDR engine across straight walks, loop returns,
zig-zag paths, variable speeds, and natural obstacle-avoidance paths, yielding a mean
drift of 0.40\,m and 97.5\% step precision across an average of 16.4/16.6 detected
steps per trial (range: 0.18--0.85\,m drift, 94.4--99.1\% precision), with the highest
drift recorded on the return-loop trial due to accumulated yaw error during repeated
turns. The mean drift satisfies the room-level accuracy
requirement, guiding users to within arm's reach of a lost item, and motivates future
integration of lightweight SLAM for multi-room drift correction.

An $N=3$ evaluation confirmed PDR generalizability; tuning the step threshold to 1.08\,G captured cautious, elderly-typical shuffling with $\le$2 steps variance per trial. A table-rest heuristic paused tracking within 1.5\,s, accumulating zero false steps during removal. A 5\,m unstructured loop (with swaying and 180$^\circ$ turns) yielded 0.60\,m drift. Stress testing revealed a wireless failure boundary: rapid rotations caused packet batching and gyroscope under-integration (4.5\,m drift). Applying a dynamic time-averaged integration method to mitigate Wi-Fi jitter, alongside a 1.62 empirical yaw multiplier, reduced this drift to 0.83\,m. Remaining errors are corrected by visual optical-flow anchors.

\subsection{Object Detection Performance}
The YOLOv11n model achieves mAP@0.5 of \textbf{0.874} after fine-tuning on the
target nine-class item hierarchy using mosaic augmentation.
A near-perfect AP of 0.931 for the ``octopus card'' class is especially significant for
daily independence use cases. The operational confidence threshold is set at 0.50,
prioritizing precision over recall to ensure high spatial certainty before notifying
users.
For micro-objects ($<$5\,cm$^2$, e.g., keys, earphones), recall was 53.3\% at 1.5\,m, degrading to 6.7\% at 3.0\,m due to edge-sensor resolution limits (OV3660 3\,MP sensor over Wi-Fi). The Siamese module achieved 100\% (5/5) discrimination accuracy on visually distinct items and a 100\% (5/5) intentional generalization rate on visually identical distractors, maximizing spatial recall probabilities.

\subsection{System Latency and Efficiency}
The edge-computing architecture achieves a total system latency of 36.7\,ms, resulting
in a negligible 4.4\,cm positional error at standard walking
speed. End-to-end voice query latency is 2--3 seconds. The
0.5\,m Distance Filter reduces database write frequency by 70\%, enabling extended
operation without maintenance. Table~\ref{tab:system} summarises key benchmarks.

\begin{table}[t]
\caption{System Performance Summary}
\label{tab:system}
\begin{tabular}{lc}
\toprule
\textbf{Metric}                   & \textbf{Value}     \\
\midrule
YOLO mAP@0.5                      & 0.707              \\
YOLO inference time               & $<$15\,ms/frame    \\
Edge-to-server latency            & $<$100\,ms         \\
Total system latency              & 36.7\,ms           \\
Voice query latency               & 2--3\,s            \\
Mean PDR drift (5 trials)         & 0.399\,m           \\
Step precision                    & 97.50\%            \\
Privacy Shield success rate       & 100\%              \\
DB write reduction (0.5m filter)  & 70\%               \\
\bottomrule
\end{tabular}
\end{table}

\subsection{User Perception Study}

\begin{figure}[t]
  \centering
  \includegraphics[width=0.8\linewidth]{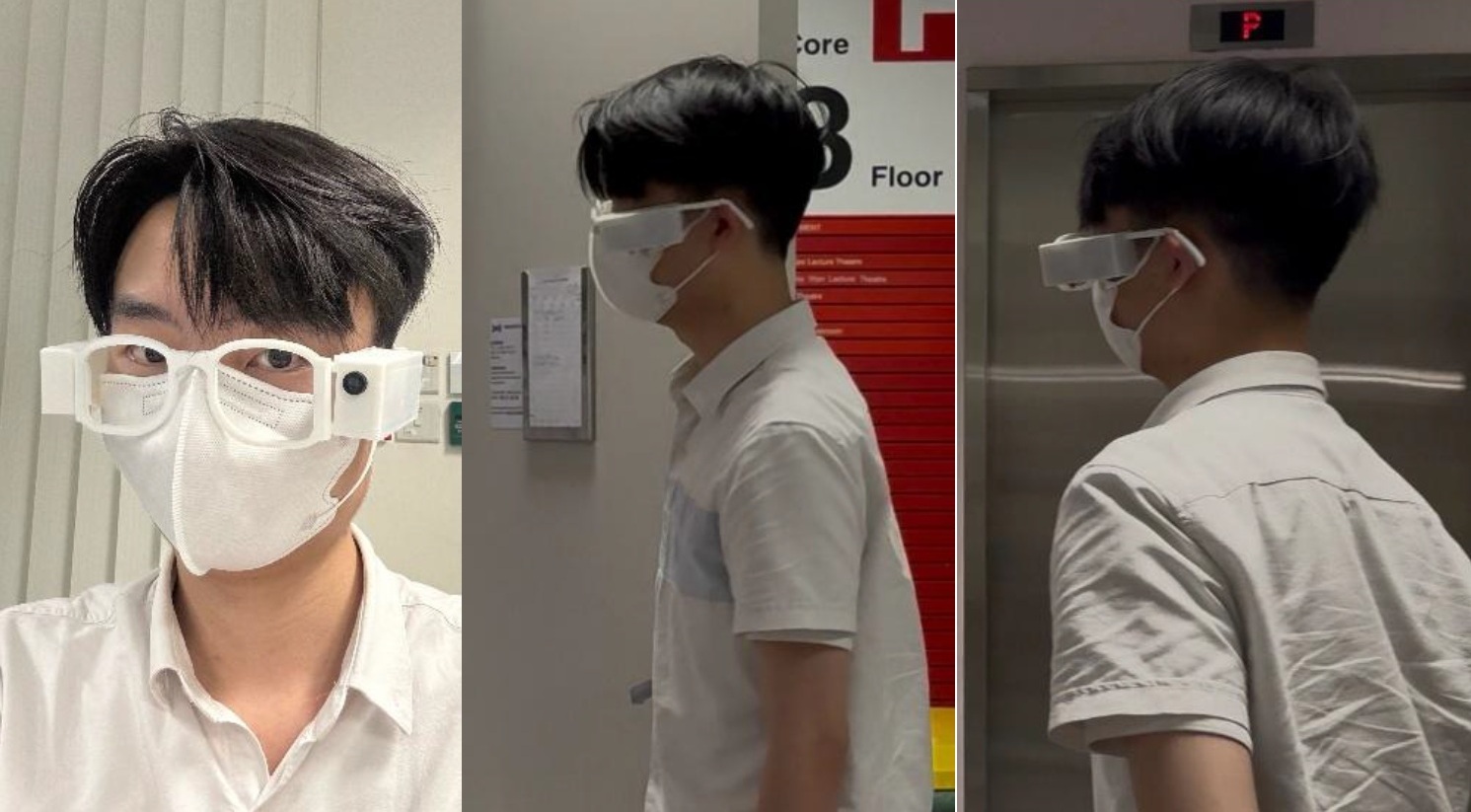}
  \caption{A study participant wearing the deployed ItemCompass eyewear prototype during the usability evaluation.}
  \Description{Photograph of a study participant wearing the 3D-printed
               eyewear prototype housing the ESP32-S3 camera and IMU
               module, used during the ISO 9241-11 usability
               evaluation.}
  \label{fig:userwearing}
\end{figure}

A usability evaluation in accordance with ISO~9241-11 was conducted with $N=31$ participants. The Accuracy \& Trust dimension scored 4.6/5.0, affirming strong reliability. Comfort scored 3.5/5.0, indicating miniaturization as the key next iteration. The Privacy Shield achieved a 100\% autonomous detection rate for toilet zones, confirming ethical guardrails do not degrade performance.

\section{Design Implications for Wearable Interaction}
ItemCompass surfaces three design principles relevant to the workshop theme. By
running detection and VIO continuously, it treats live movement through the home as
an ongoing, automatically indexed record; this ``digital twin'' paradigm could ground
wearable games in which physical domestic space accumulates narrative history, object
traces, and event logs, creating a personalized persistent-world experience without
game-specific infrastructure. The 2--3 second voice query loop demonstrates that full
LLM reasoning can be delivered locally at a conversational pace without the cloud
dependency, opening design space for voice-driven AR games, guided exploration
experiences, and accessibility tools. Finally, the 100\% success rate of the
hardware-level Privacy Shield demonstrates that ethical guardrails can be enforced at
the sensor edge rather than delegated to cloud policy. A pattern essential for
wearables intended for domestic, care, and entertainment contexts, where user dignity
and bystander consent are equally important, and one that directly addresses the
workshop's focus on accessibility and well-being for vulnerable populations.

\section{Limitations and Future Work}
Three challenges remain. First, PDR drift accumulates in multi-room trajectories
with repeated turns. Future work will integrate lightweight SLAM for long-range
heading correction. Second, micro-object detection (e.g., AirPods) is limited by
the hardware's optical resolution. The Siamese re-identification module also
discriminates well between visually distinct objects, but cannot disambiguate
identical items owned by different household members. Third, the chassis comfort
score of 3.5/5.0 calls for a miniaturized, PCB-integrated frame. Future versions
will also explore multi-user shared spatial memory for collaborative item tracking.

\section{Conclusion}
ItemCompass demonstrates that persistent, voice-queryable spatial memory of a domestic
environment can be delivered through a fully wearable, tag-free edge-AI system running
on commodity hardware, without cloud dependency or home modification. A mean
localization drift of 0.40\,m, 97.5\% step precision, and mAP@0.5 of 0.874 confirm
both spatial accuracy and detection reliability. High user trust scores and a 100\%
privacy-shield success rate validates the system's readiness for real-world deployment.
Beyond its assistive application, ItemCompass illustrates how always-on wearable
vision can serve as a design medium for ambient, conversational, and spatially
grounded interaction---principles directly relevant to the design of wearable
experiences for games, entertainment, and everyday ubiquitous computing.

\section*{Ethics and Privacy Statement}
ItemCompass operates within the home and continuously captures first-person visual
data. The system retains only object class labels, spatial coordinates, and headings;
no raw images are persisted beyond inference. All processing is local---no data leaves
the home network. The Privacy Shield autonomously suspends logging in sensitive zones
(bathrooms/toilets) with a 100\% detection success rate in evaluation. The $N=31$
usability study and $N=3$ PDR evaluations were conducted with informed consent, and
participants retained control over their data retention and deletion. Designers
adapting this system for games or entertainment must consider additional safeguards
when incidental bystander capture is possible.

\begin{acks}
The authors thank the PolyU Industrial Centre for technical support in hardware
prototyping and chassis fabrication. Some illustrative icons in
Figure~\ref{fig:teaser} were generated with AI assistance; figure composition and
technical content are the authors' own.
\end{acks}

\bibliographystyle{ACM-Reference-Format}
\bibliography{reference}

\end{document}